\documentclass[12pt]{article}
\usepackage{amsmath}
\usepackage{amsfonts}
\usepackage{amssymb}
\usepackage{graphicx}

\usepackage{color}
\usepackage{ulem}

\definecolor{mycolor}{rgb}{1,0.5,0.}

\newcommand{\x}{{\vec x}}

\newcommand{\Eq}[1]{Eq.~(\ref{#1})}
\newcommand{\ave}[1]{\langle{#1}\rangle}

\def\beq{\begin{equation}}
\def\eeq#1{\label{#1}\end{equation}}
\def\eeqn{\end{equation}}

\def\beqa{\begin{eqnarray}}
\def\eeqa#1{\label{#1}\end{eqnarray}}
\def\eeqan{\end{eqnarray}}

\let\bar=\overbar

\def\Dslash{\not{\hbox{\kern-4pt $D$}}}
\def\dslash{\not{\hbox{\kern-2pt $\del$}}}

\def\msb{{\bar{\ssstyle M \kern -1pt S}}}

\usepackage{fancyhdr,graphicx}
\def\Title#1{\begin{center} {\Large {\bf #1} } \end{center}}

\begin{document}

\Title{Inhomogeneous chiral symmetry breaking phases in isospin-asymmetric matter}

\bigskip\bigskip

\begin{raggedright}

{\it 
Daniel Nowakowski$^{1}$\!,~Michael Buballa$^{1}$\!,~Stefano Carignano$^{2}$\!,~and Jochen Wambach$^{1,3}$\\
\bigskip
 $^1$ Theoriezentrum, Institut f\"ur Kernphysik, TU Darmstadt, Schlossgartenstra\ss{}e 2, 64289 Darmstadt, Germany \\
\bigskip
 $^2$ Department of Physics, The University of Texas at El Paso, 500 West University Avenue, El Paso, Texas 79968, United States of America\\
\bigskip
$^3$ GSI Helmholtzzentrum f\"ur Schwerionenforschung GmbH, Planckstra\ss{}e 1, 64291 Darmstadt, Germany
}

\end{raggedright}

\begin{abstract} 
We investigate the effects of isospin asymmetry on inhomogeneous chiral symmetry breaking phases within the two-flavor NJL model. After introducing a plane-wave ansatz for each quark-flavor condensate, we find that, as long as their periodicities are enforced to be equal, a non-zero isospin chemical potential shrinks the size of the inhomogeneous phase. The asymmetry reached in charge neutral matter is nevertheless not excessively large, so that an inhomogeneous window is still present in the phase diagram. Lifting the constraint of equal periodicities alters the picture significantly, as the inhomogeneous phase survives in a much larger region of the phase diagram.
\end{abstract}

\section{Introduction}

Chiral symmetry, which is an approximate symmetry of the QCD Lagrangian in the sector of up and down quarks, 
is spontaneously broken in vacuum by the appearance of a non-zero chiral condensate $\ave{\bar\psi\psi}$.
It is known from lattice calculations that, for vanishing chemical potential $\mu$, chiral symmetry gets (approximately) restored 
at high temperatures $T$ in a crossover transition taking place at $T \approx 150$~MeV.
At low temperature and $\mu \neq 0$ on the other hand, where standard lattice methods are not applicable,
many model calculations, performed e.g., within the Nambu--Jona-Lasinio (NJL) model~\cite{Asakawa:1989bq}
 or the quark-meson model~\cite{Scavenius:2000qd}, suggest that there is a first-order phase transition where  
the chiral condensate discontinuously drops from a large value to almost zero.

Most of these studies have been performed under the tacit assumption that $\ave{\bar\psi\psi}$ is spatially constant, a condition which might turn out to be too restrictive.
Although inhomogeneous phases with a spatially varying chiral order parameter have been discussed already 25 years ago~\cite{Kutschera:1990xk} (see also \cite{Broniowski:2011ef}),  this possibility has gained increased attention only recently, 
after several new studies have confirmed its relevance, 
see~\cite{Buballa:2014tba} for a review. 
In particular in the NJL model one finds that at low temperatures an ``inhomogeneous island'' appears between the homogeneous phases with broken and restored chiral symmetry.

Even if its effect on the equation of state is rather small \cite{Carignano:2015kda},
inhomogeneous quark matter could have interesting consequences for the physics of compact stars, in particular for their transport and cooling properties.
So far, however, most studies of inhomogeneous chiral condensates have been performed for isospin symmetric matter, whereas it is known that the constraints of electric neutrality and beta equilibrium lead to a considerable isospin asymmetry in compact stars.
In order to implement a more realistic scenario, we have therefore investigated the influence of a non-vanishing isospin chemical potential $\mu_I = \mu_u - \mu_d$
on the inhomogeneous phase within an NJL model.
In these proceedings we present some first results of this study, while
a more detailed discussion will be given elsewhere~\cite{NBCW}.

\section{Model}
\label{sec:model}

We employ the two-flavor NJL model, defined by the Lagrangian~\cite{NJL2}
  \begin{align}
  \mathcal{L} = \bar{\psi} \left(i\gamma^\mu \partial_\mu-\hat{m}\right)\psi +  G \left(\left(\bar{\psi}\psi\right)^2+\left(\bar{\psi}i\gamma_5\tau^a\psi\right)^2\right),
\label{eq:LNJL}
  \end{align}
where $\psi=(u,d)^T$ is the $4N_cN_f$-dimensional quark spinor for $N_f=2$ flavors and $N_c=3$ colors, and $G$ denotes a dimensionful coupling. Here, $\gamma^\mu$ are the Dirac matrices, $\tau^a$ the Pauli matrices acting in flavor space, and $\hat{m}=\text{diag}(m_u,m_d)$ is the current mass matrix. 

To determine the thermodynamically favored ground state of the system, we evaluate the grand potential per unit volume,
generalizing the formalism described e.g.~in Ref.~\cite{Buballa:2014tba} to non-zero isospin chemical potential
(see~\cite{NBCW} for details).
To this end we employ the mean-field approximation in the presence of the flavor diagonal\footnote{By restricting 
the analysis to flavor diagonal condensates we neglect the possibility of charged pion condensation.
Therefore our analysis is only valid for $|\mu_I| < m_\pi$, which is nevertheless sufficient for our purpose.
} scalar and pseudoscalar 
condensates 
 \begin{align*}
     \begin{array}{lll}
       S_f(\x)=\left\langle \bar{f} f\right\rangle, & \quad\qquad & P_f(\x)=\left\langle\bar{f} i\gamma_5f\right\rangle,
     \end{array}
   \end{align*}
with $  f\in\{u,d\}$.     
Here an explicit spatial dependence of the condensates is retained, since we want to analyze the emergence of inhomogeneous chiral symmetry breaking in our model.

In order to keep the problem tractable we restrict ourselves to simple one-dimensional spatial modulations. 
Specifically, for each flavor $f$ we consider the ansatz 
 \begin{align}
  S_f(z)=-\frac{\Delta_f}{4G} \cos{(q_f z)} \,,\quad\qquad P_f(z)=-\frac{\Delta_f}{4G} \sin{(q_f z)}\,,\label{eq:condensates}
 \end{align}
with amplitudes $\Delta_f$ and wave numbers $q_f$.
 Moreover we require that the ratio $R = q_u/q_d$ is a rational number, so that the system is overall periodic,
 characterized by a period length $L$. 
The mean-field thermodynamic potential can then be written as
\begin{align}
  \Omega(T,\{\mu_f\};\{\Delta_f\},\{q_f\})   = \sum_{f=u,d}\Omega_{\text{kin}}^f +\Omega_\text{cond}+\text{const.},
\label{eq:thermpot}
\end{align}
with
\begin{align*}
  \Omega_\text{cond}=\frac{G}{L}\int_0^{L} dz\, \Big(\big(S_u(z)+S_d(z)\big)^2 +  \big(P_u(z)- P_d(z)\big)^2\Big),
\end{align*}
and
\begin{align*}
 \Omega_{\text{kin}}^f=\sum_{\lambda}\Bigg[& E^f_\lambda+T\log{\bigg(1+\exp{\bigg(-\frac{E^f_\lambda-\mu_f}{T}\bigg)}\bigg)}+T\log{\bigg(1+\exp{\bigg(-\frac{E^f_\lambda+\mu_f}{T}\bigg)}\bigg)}\Bigg],\nonumber
\end{align*}
where $\{E_\lambda^f\}$ are the eigenvalues of the effective mean-field Hamiltonian
\begin{align}
 H_f=-i\gamma^0\gamma^i\partial_i+\gamma^0 \big[m_f-2G\big(&\Delta_f \cos{(q_f z)}+\Delta_h \cos{(q_h z)}\label{eq:hamiltonian}\\ & +i\gamma^5 (\Delta_f \sin{(q_f z)}-\Delta_h \sin{(q_h z)})\big)\big] \nonumber,
\end{align}
and $h\in\{u,d\}$ with $f\neq h$.

The problem of calculating the thermodynamic potential is then essentially reduced to the determination of the eigenvalue spectrum of $H_f$.
This will be done numerically, after performing a Fourier transform to momentum space.

Since the NJL model is non-renormalizable, we need to regularize the divergent contributions in the thermodynamic potential. For this, a Pauli-Villars regularization scheme is applied to the diverging vacuum part of the grand potential (see \cite{Nickel:2009wj} for details).

At this stage we are then able to minimize the thermodynamic potential at given $(T,\mu_u,\mu_d$) with respect to the variational parameters $\{\Delta_f\}$ and $\{{q}_f\}$, to determine the energetically favored ground state.

\section{Phase structure}

In this section we discuss our numerical results for the phase diagram in isospin-asymmetric matter,
focusing on the size and the properties of the inhomogeneous phase.
For this we introduce the average chemical potential $\bar{\mu}=(\mu_u+\mu_d)/2$
and present phase diagrams in the $\bar{\mu}-T$ plane for various values of $\mu_I$.
The individual flavor chemical potentials are then given by 
\begin{align*}
 \mu_{u,d}=\bar{\mu}\pm\frac{\mu_I}{2},
\end{align*}
from which it is obvious that changing the sign of $\mu_I$ only interchanges the roles of the up and down quarks. 
The shape of the phase diagrams therefore only depends on the modulus of $\mu_I$. 

For simplicity we restrict our calculations to the chiral limit, $m_f=0$.\footnote{
We are aware that charged pions condense as soon as $\mu_I > m_\pi$, which in the chiral limit would correspond to $\mu_I > 0$. 
Here, however, we regard $m_f = 0$ only as a useful approximation to evaluate the thermodynamic potential,
and assume that charged pion condensation does not occur below the physical pion mass.
}
The remaining model parameters, the coupling $G$ and the Pauli-Villars cutoff $\Lambda$, are determined by reproducing the pion decay constant in the chiral limit $f_\pi=88\,\text{MeV}$ and a constituent quark mass of $300\,\text{MeV}$ in vacuum.

\subsection{Equal periodicities}
\label{eq:equal}

\begin{figure}[t]
  \centering
  \includegraphics[width=0.48\textwidth]{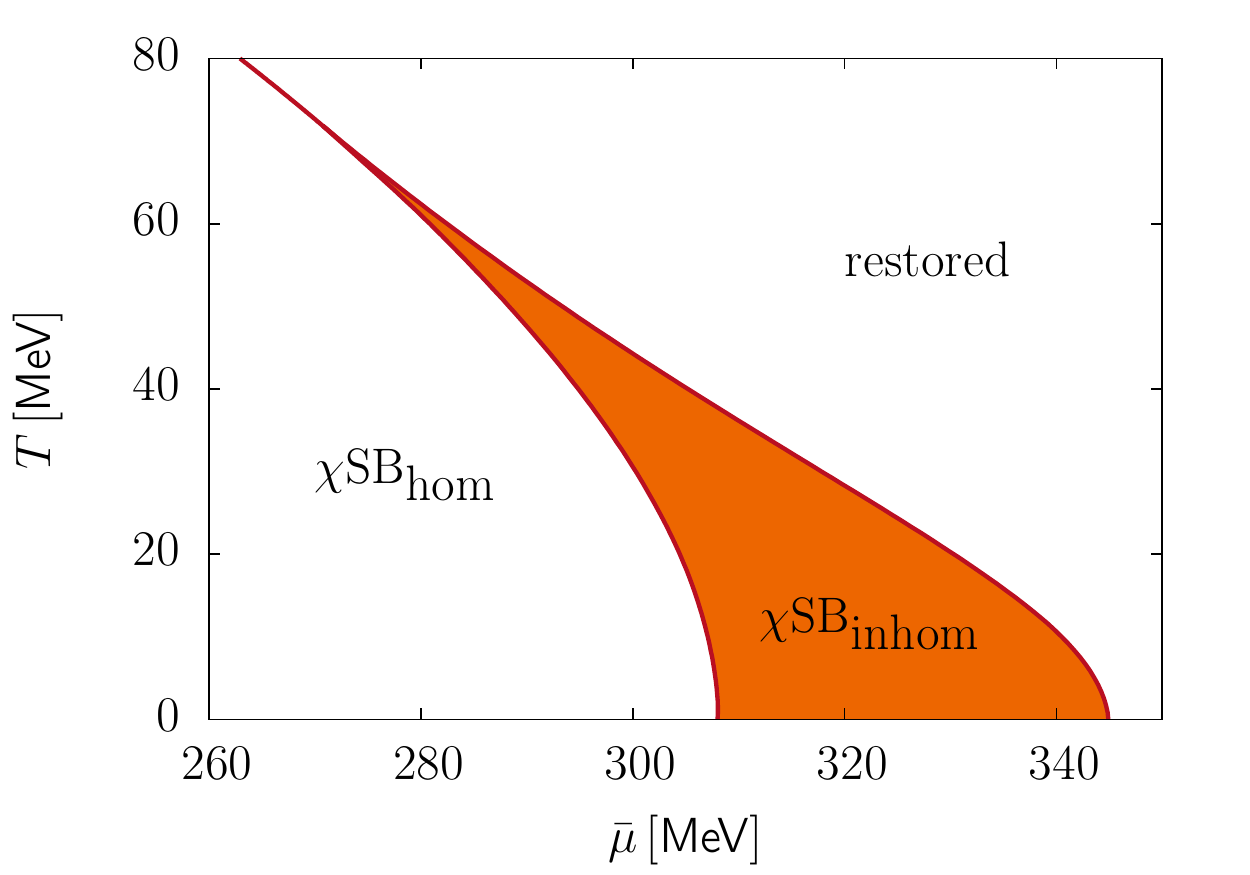}
  \caption{Phase diagram in the $\bar{\mu}-T$ plane for $\mu_I=0$. The shaded area indicates the region where the inhomogeneous solution with a plane-wave modulation is favored over a homogeneous solution. This ``inhomogeneous island'' is located between a region with homogeneously broken chiral symmetry ($\chi\text{SB}_\text{hom}$) on the left and an area where chiral symmetry is restored on the right.}
 \label{fig:phased0}
\end{figure}

To have a well-defined starting point for our investigation, 
we show in Fig.~\ref{fig:phased0} the phase diagram of the isospin-symmetric case, 
$\mu_I = 0$.
The region where the inhomogeneous phase is favored is indicated by the shaded area. 
Because of the isospin symmetry, the ansatz  \Eq{eq:condensates} 
reduces to the so-called (dual) chiral density wave (CDW)~\cite{Kutschera:1990xk,Nakano:2004cd} 
where the amplitudes and wave numbers of both flavors are equal.
Strictly speaking we have $\Delta_u = \Delta_d$ and $q_u = -q_d$ where the minus sign arises as a consequence of
the isovector nature of the pseudoscalar interaction in \Eq{eq:LNJL}. 

Next, we turn to $\mu_I \neq 0$.
As a first step, we restrict \Eq{eq:condensates} to a CDW-like ansatz with arbitrary amplitudes but
equal periodicities, $q_u = -q_d$.
This has the advantage that the Hamiltonian $H_f$ can be diagonalized analytically, considerably simplifying the problem.

\begin{figure}[t]
  \centering
\begin{minipage}{0.32\textwidth}
 \includegraphics[width=0.99\textwidth]{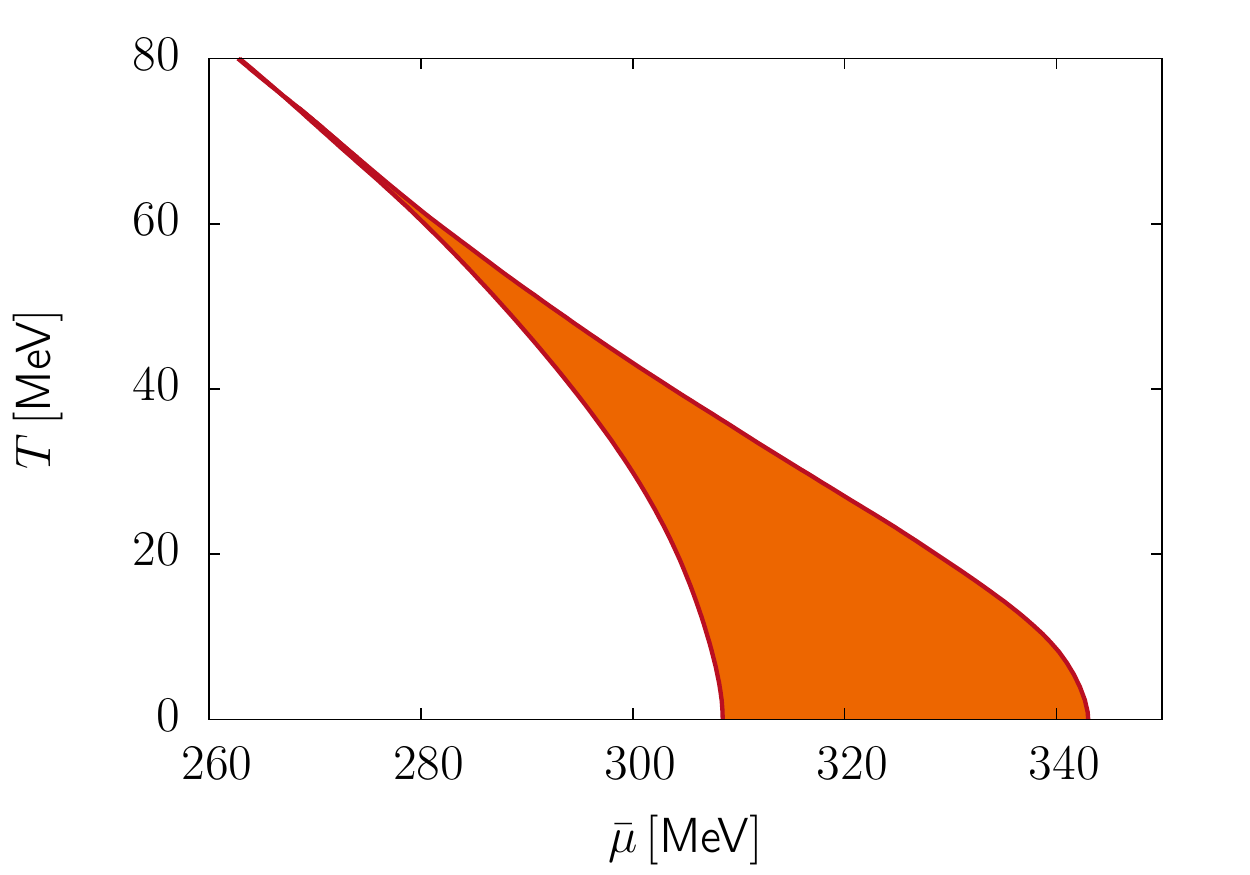}
\begin{center}
 $|\mu_I|=20\,\text{MeV}$
\end{center}
\end{minipage}
\begin{minipage}{0.32\textwidth}
 \includegraphics[width=0.99\textwidth]{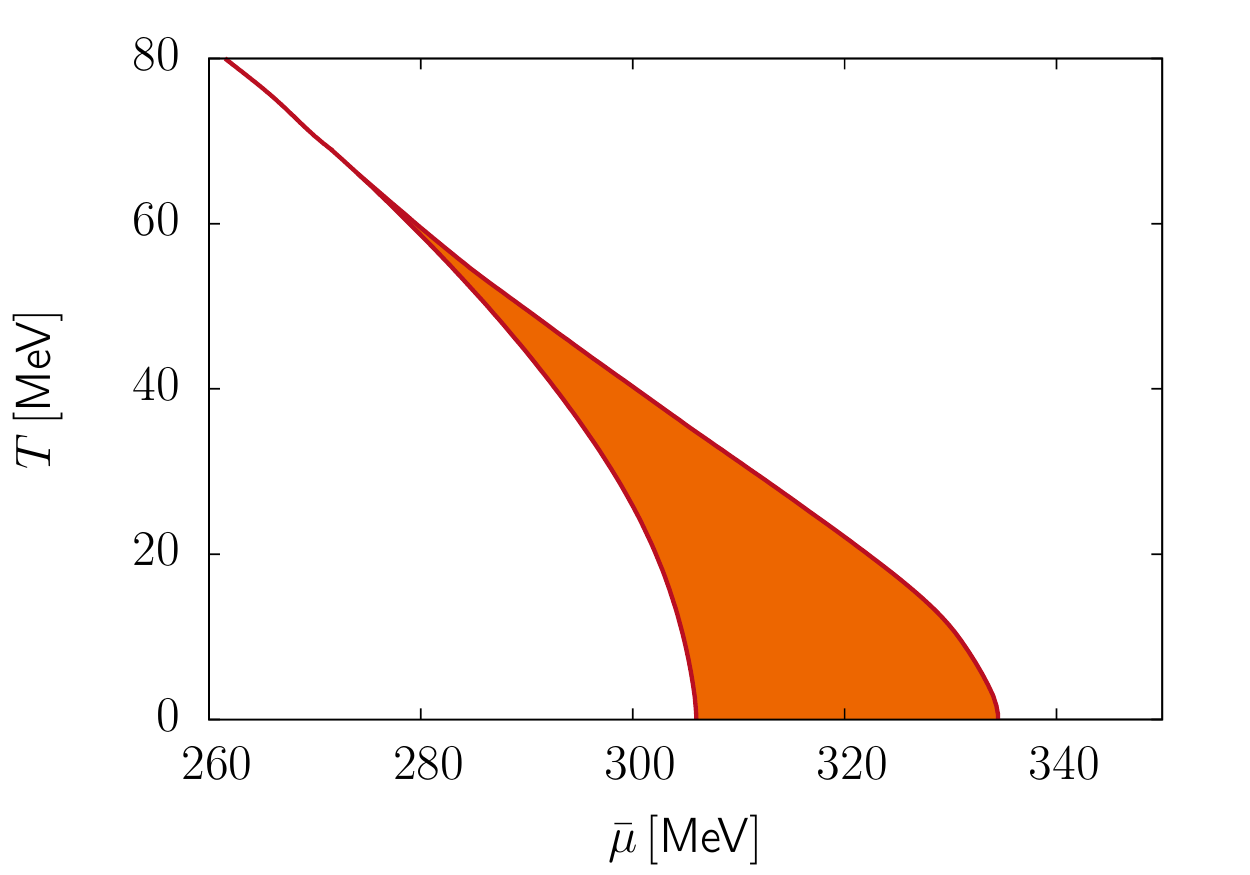}
\begin{center}
 $|\mu_I|=60\,\text{MeV}$
\end{center}
\end{minipage}
\begin{minipage}{0.32\textwidth}
 \includegraphics[width=0.99\textwidth]{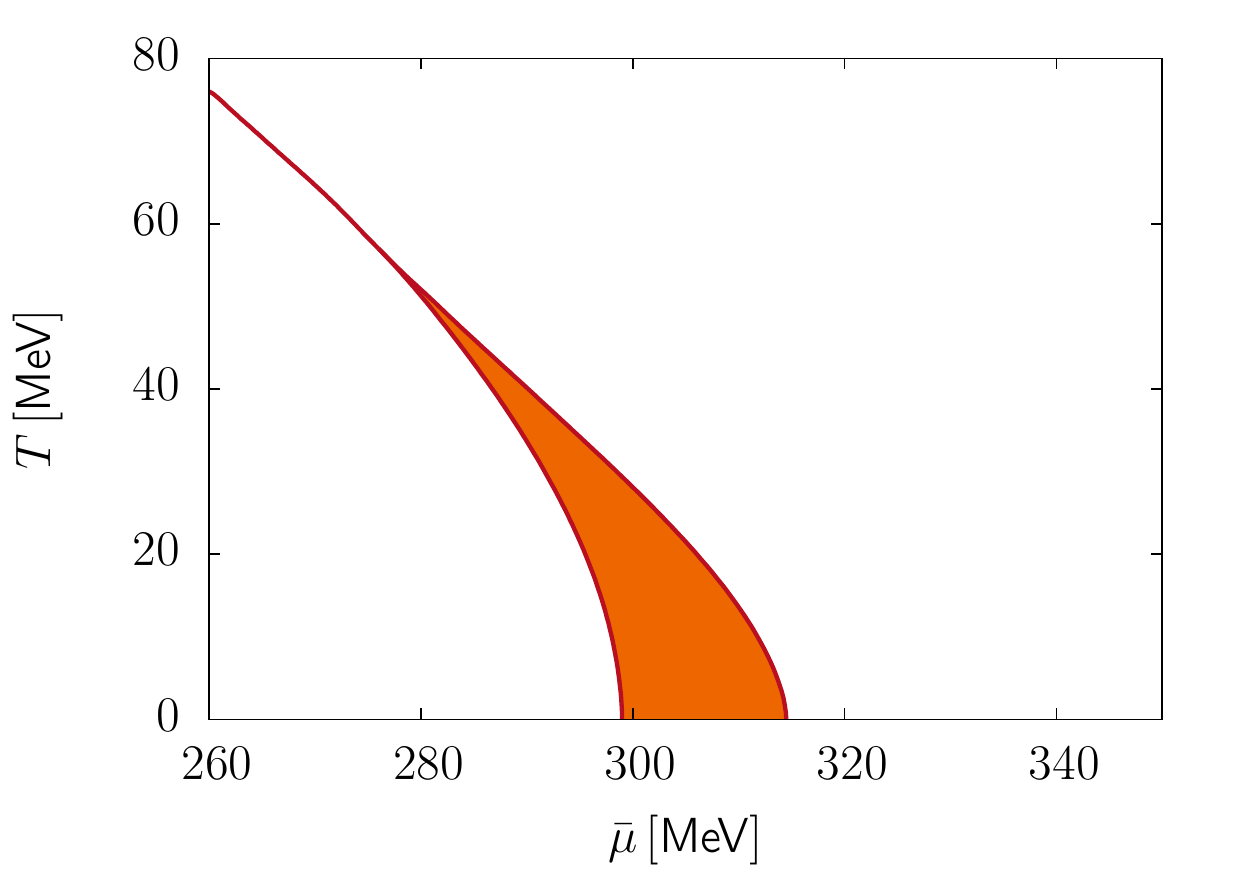}
\begin{center}
 $|\mu_I|=120\,\text{MeV}$
\end{center}
\end{minipage}
  \caption{\label{fig:phased}Phase diagram in the $\bar{\mu}-T$ plane for three different values of $\mu_I$. The shaded areas indicate the regions where a CDW-like modulation of the condensates with $q_u = -q_d$ is favored over a homogeneous solution.}
 \end{figure}

The resulting phase diagrams for three different values of $|\mu_I|$ are displayed in  Fig.~\ref{fig:phased}.
We find that for this CDW-type ansatz the inhomogeneous phase shrinks as $\mu_I$ increases and its onset moves to lower values of $\bar\mu$. The inhomogeneous window nevertheless appears to be relatively robust, surviving beyond $|\mu_I| = 120$~MeV.

It is very plausible that the observed reduction is at least partially due to the fact that 
our assumption of equal periodicities is too restrictive:
For isospin-symmetric matter it is known that the wave number of the CDW strongly depends on the chemical potential,
roughly being of the order of $2\mu$.
One should therefore expect that for $\mu_u \neq \mu_d$ up and down quarks would favor different periodicities,
although the situation is complicated by the fact that up- and down-quark condensates mix in the Hamiltonian,   
see  \Eq{eq:hamiltonian}.
We will come back to this issue in Section~\ref{sec:unequal}.

\subsection{Electric charge neutrality}

In compact stars, the isospin chemical potential is no longer an independent external parameter, but is fixed 
by the requirement of global electric charge neutrality and beta equilibrium. 
In order to describe this situation a leptonic component must be added to our model.

For this we consider an ideal gas of massless electrons and assume that neutrinos can freely leave the star. 
The system can then be characterized by two chemical potentials $\mu$ and $\mu_Q$,
corresponding to two conserved  quantities: the net quark number and the electric charge.
Accordingly, the chemical potentials of the quarks and electrons are given by
\begin{align}
 \mu_u=\mu+\frac{2}{3}\mu_Q,\qquad \mu_d=\mu-\frac{1}{3}\mu_Q\label{eq:chargeneutral}, \qquad
 \mu_e = -\mu_Q ,
\end{align}
and thus $\mu_I = \mu_Q$ (while $\bar\mu = \mu + \mu_Q/6 \neq \mu$).
With this, we can write the total thermodynamic potential as
\begin{align}
  \Omega_\text{tot}(T,\mu,\mu_Q)
  =
  \Omega(T,\{\mu_f\})  + \Omega_e(T,\mu_e) \,,
\end{align}
with the quark contribution $\Omega$  given by \Eq{eq:thermpot}
and the electron contribution $\Omega_e$.
The requirement of global electric charge neutrality then takes the form
\begin{align}
n_Q = 
-\frac{\partial \Omega_\text{tot}}{\partial \mu_Q}
=
\frac{2}{3} n_u -\frac{1}{3} n_d -n_e \stackrel{!}{=}0,
\label{eq:chargeneutral1}
\end{align}
where $n_f=-\partial\Omega/\partial\mu_f$ are the (spatially averaged) quark number densities and $n_e = -\partial\Omega_e/\partial\mu_e$ denotes the electron number density.

\begin{figure}[t]
  \centering
  \includegraphics[width=0.48\textwidth]{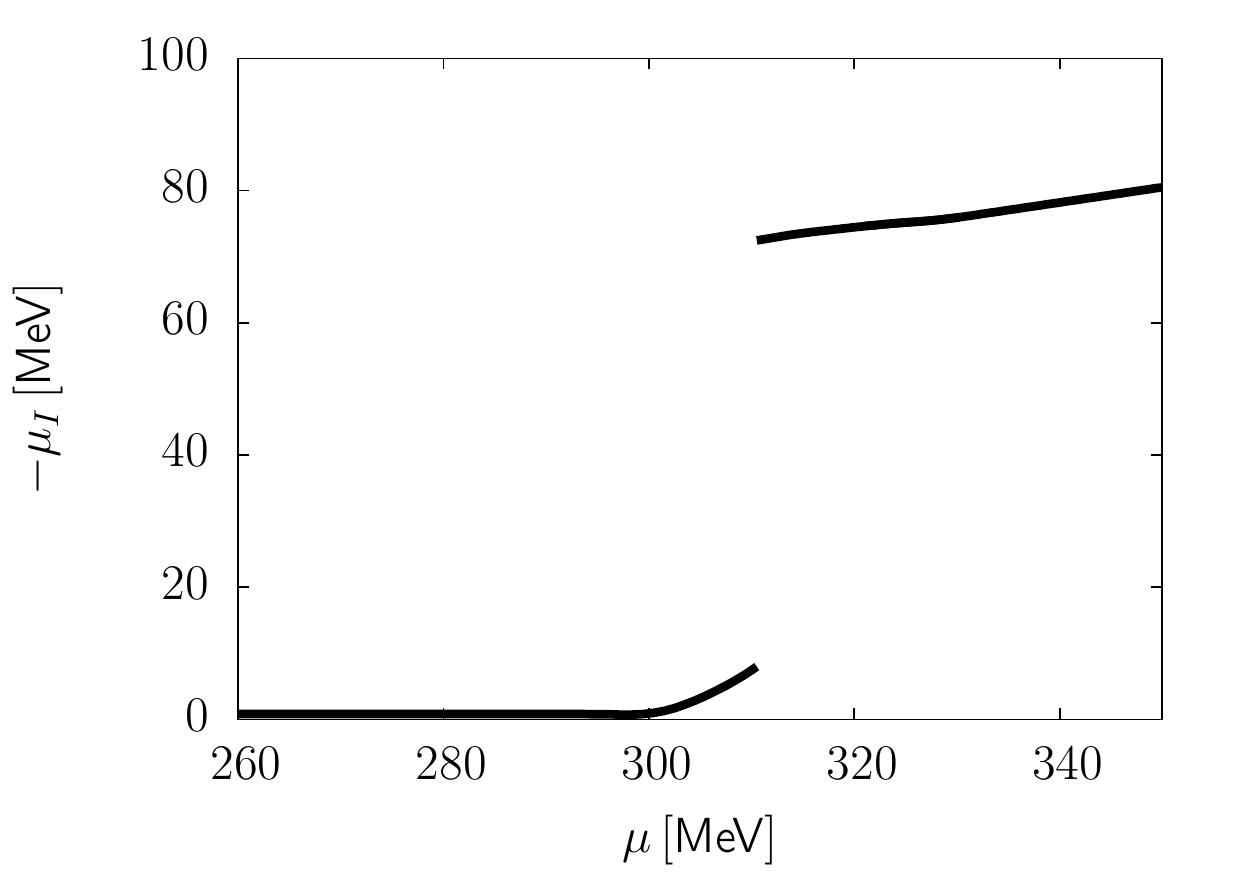}
 \caption{\label{fig:chargeneutral1}
 Isospin chemical potential $\mu_I \equiv \mu_Q$ as a function of the quark-number chemical potential $\mu$
 for charge-neutral matter in beta equilibrium 
 at $T=1\,\text{MeV}$, allowing for CDW-like chiral condensates with $q_u=-q_d$.}
\end{figure}

\begin{figure}[ht]
  \centering
  \includegraphics[width=0.48\textwidth]{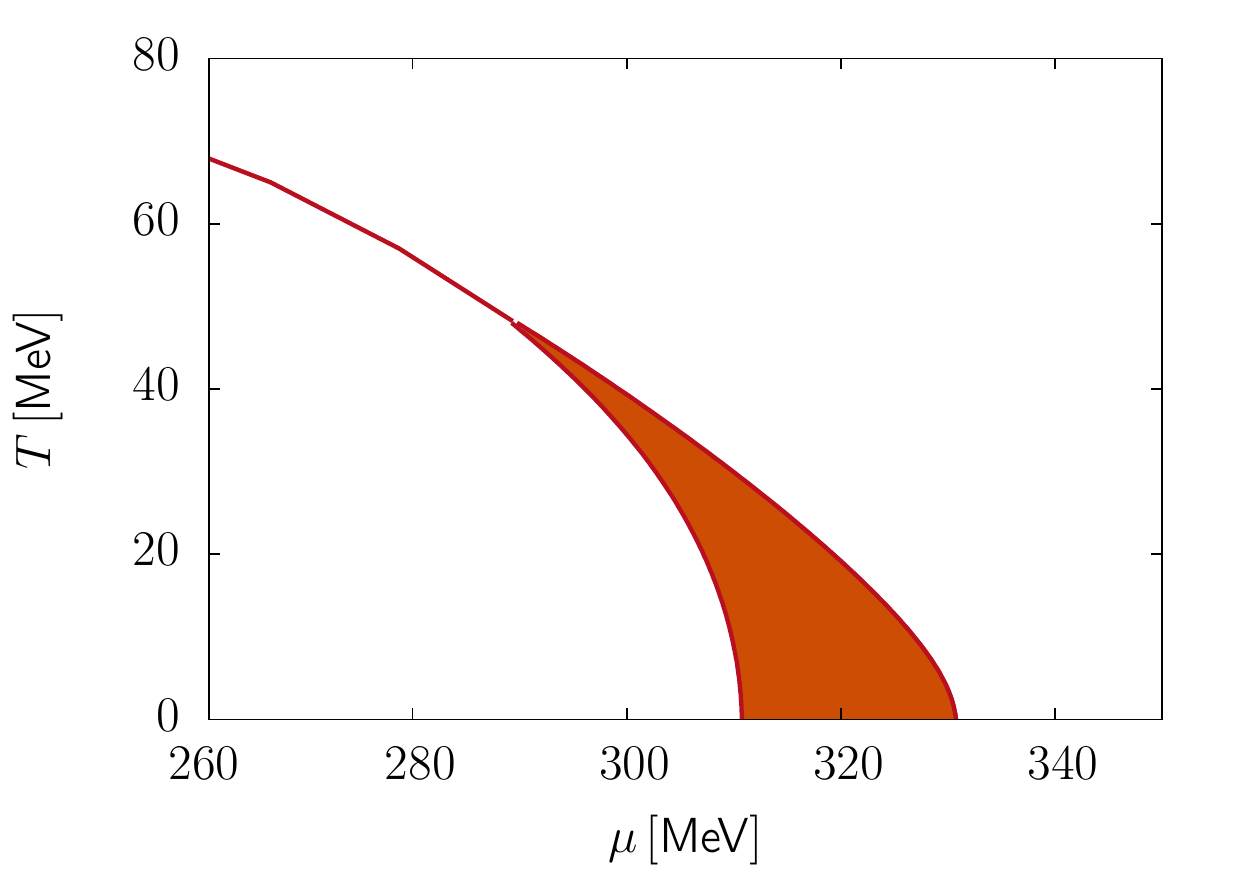}
\caption{\label{fig:chargeneutral2}Phase diagram in the $\mu-T$-plane for electrically neutral matter in beta equilibrium.
The shaded area indicates the region where a CDW-like modulation of the chiral condensates with $q_u = - q_d$ is energetically favored over a homogeneous solution.}
\end{figure}

Simultaneously to this condition, we have to minimize the thermodynamic potential with respect to the amplitudes $\Delta_f$
and wave numbers $q_f$ of the chiral condensates. 
Thereby we again restrict ourselves to the simplified case where we enforce $q_u = -q_d$.
The resulting value of $\mu_Q$ at zero temperature\footnote{To be precise, the calculations have been performed for
$T=1$~MeV.
}
as a function of $\mu$ is depicted in Fig.\;\ref{fig:chargeneutral1}. 
Since $\mu_Q$ turns out to be always below 80 MeV in the region where the inhomogeneous phase is expected to appear, in light of the results of Section~\ref{eq:equal} we expect the latter to be still present, albeit decreased in size. This is exactly what we observe, as shown in Fig.\;\ref{fig:chargeneutral2}.
On the other hand, as previously mentioned, the inhomogeneous region might grow again if the artificial constraint $q_u = -q_d$ is lifted.

\subsection{Unequal periodicities}
\label{sec:unequal}

Finally, we therefore relax the artificial constraint $q_u = -q_d$ and allow for $|R| = |q_u/q_d| \neq 1$.\footnote{
The relative sign between $q_u$ and $q_d$ is only relevant for equal periodicities, i.e.,  for $R=\pm 1$. 
We can therefore choose $R$ to be positive in all other cases.
}
As explained in Section~\ref{sec:model},
we require an overall periodicity of the system, which implies that $R$ must be a rational number.
Since any real number can be approximated to arbitrary accuracy by rational numbers, this is not a severe limitation in principle. 
In practice, however, we are only able to  investigate a rather small number of ratios.

Indeed, for CDW-like modulations with unequal periodicities
for up- and down-quark condensates, the mean-field Hamiltonian $H_f$ cannot be diagonalized analytically any more and we have 
to resort to determine the eigenvalue spectrum numerically. 
This turns out to be rather time consuming, so that we restrict ourselves to a small set of ratios, $R\in\{-1,4/3,2,5\}$, 
from which we determine the energetically preferred solution at given values of $T$, $\bar{\mu}$ and $\mu_I$. 

\begin{figure}[htb]
  \centering
  \includegraphics[width=0.48\textwidth]{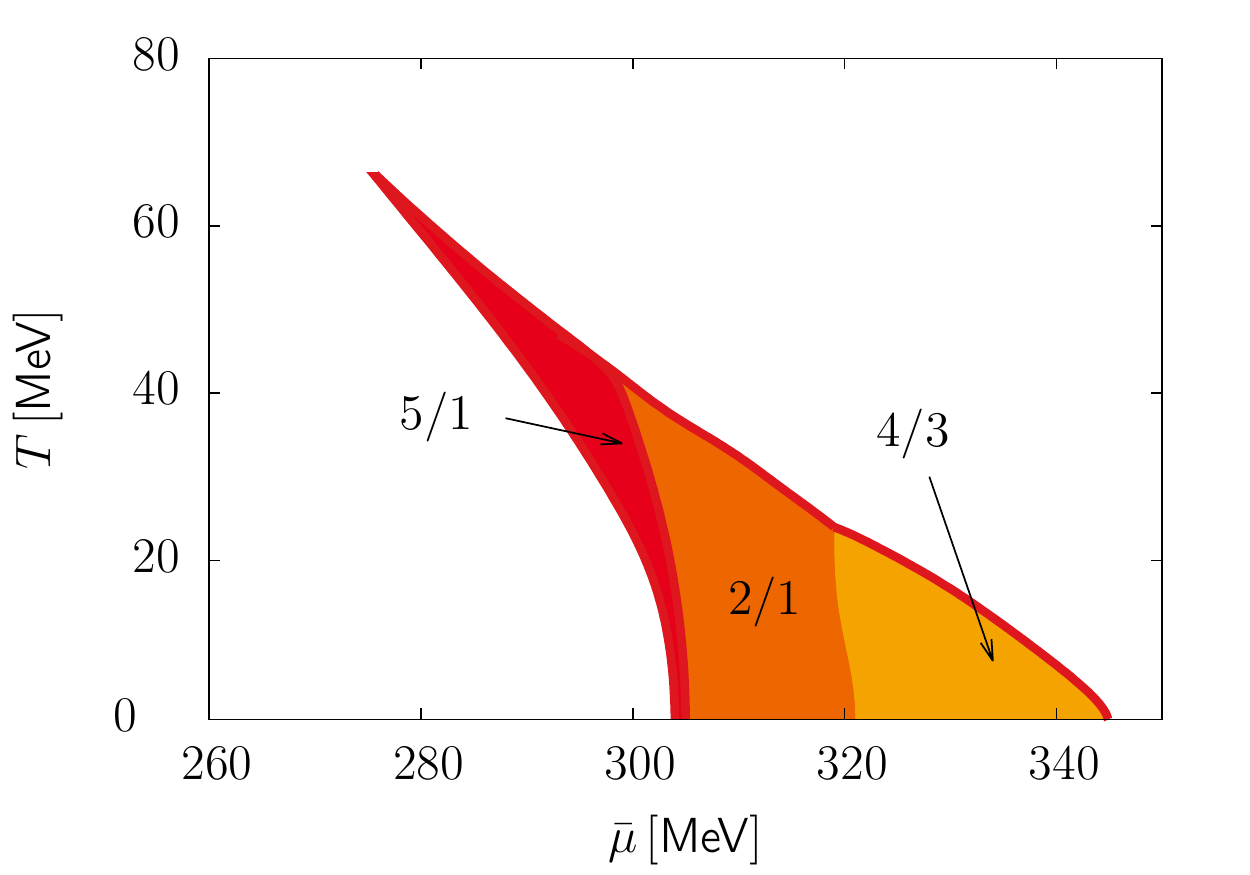}
 \caption{\label{fig:diffperiods}Phase diagram in the $\bar{\mu}-T$ plane for $\mu_I=60\,\text{MeV}$. The shaded areas indicate the
 regions where an inhomogeneous solution with the CDW-like modulation \Eq{eq:condensates} 
 and $R = q_u/q_d \in\{-1,4/3,2,5\}$ is favored over a homogeneous solution.  
 Thereby different colors correspond to different values of the energetically most favored ratio $R$, 
 as indicated by the labels.}
\end{figure}

The resulting $\bar\mu-T$ phase diagram for $\mu_I=60\,\text{MeV}$ 
is shown in Fig.\;\ref{fig:diffperiods}. We find that indeed different ratios are favored throughout different regions of the phase diagram. 
Moreover, allowing for unequal periodicities stabilizes the inhomogeneous phase considerably.
Even though we tried out only a limited number of ratios, the inhomogeneous region is now almost as large as
for isospin symmetric matter, cf.~Fig.~\ref{fig:phased0}.
 
\section{Summary}

We have investigated the effects of isospin asymmetry on inhomogeneous quark matter within the two-flavor
Nambu--Jona-Lasinio model. 
After choosing a simple plane-wave ansatz for the scalar and pseudoscalar condensates of
each flavor, we first considered a restrictive CDW-type solution where the periodicities for each flavor are forced to be equal. Since inhomogeneous particle-hole condensation is mainly a Fermi surface effect, the periodicity of each flavor is strongly related to its own chemical potential. It is then clear that, in the presence of an isospin imbalance, the requirement of equal periodicities can be very restrictive, and indeed we observed that for this ansatz the resulting inhomogeneous phase shrinks significantly as $\mu_I$ increases. After imposing charge neutrality on our system and determining self-consistently the amount of isospin imbalance, we find that the inhomogeneous window becomes smaller but does not disappear from the phase diagram.

On the other hand, if the periodicities are allowed to be unequal, the inhomogeneous phase gets stabilized against the 
isospin-imbalance effects. 
While the numerical implementation of these solutions is technically challenging, we were able to observe that, 
for a fixed $\mu_I$, the favored ratio $q_u/q_d$ 
takes rather large values near the boundary to the homogeneous chirally broken phase and
decreases with increasing $\mu$.
A more detailed discussion will be given in Ref.~\cite{NBCW}.

\subsection*{Acknowledgment}

We thank the organizers of CSQCD IV for providing a stimulating atmosphere and for financial support.
This work was also supported in part by  the Helmholtz International Center for FAIR, 
Helmholtz Graduate School for Hadron and Ion Research HGS-HIRe, 
the ExtreMe Matter Institute EMMI, and BMBF.

\end{document}